\documentclass[journal]{IEEEtran}

\usepackage{cite}
\usepackage{amsmath,amssymb,amsfonts}
\usepackage{algorithm}
\usepackage{algorithmic}
\usepackage{graphicx}
\usepackage{textcomp}
\usepackage{xcolor}
\usepackage{bm}
\usepackage{subfigure}
\usepackage{multirow}
\usepackage[normalem]{ulem}
\usepackage{tabularx}
\usepackage{xpatch}
\usepackage{soul}
\usepackage{ulem}
\usepackage{comment}

\ifCLASSINFOpdf
\else
\fi

\begin{document}

\title{Holographic MIMO Systems, Their Channel Estimation and Performance}

\author{Yuanbin~Chen,~Ying~Wang,~\IEEEmembership{Member,~IEEE,}~Zhaocheng~Wang,~\IEEEmembership{Fellow,~IEEE,}~and~Ping~Zhang,~\IEEEmembership{Fellow,~IEEE}

\thanks{This work was supported in part by the Beijing Natural Science Foundation under Grant 4222011. \textit{(Corresponding authors: Ying~Wang; Zhaocheng~Wang.)}

Yuanbin Chen, Ying Wang, and Ping Zhang are with the State Key Laboratory of Networking and Switching Technology, Beijing University of Posts and Telecommunications, Beijing 100876, China (e-mail: chen\_yuanbin@163.com; wangying@bupt.edu.cn; pzhang@bupt.edu.cn).

Zhaocheng Wang is with the Beijing National Research Center for Information Science and Technology, Department of Electronic
Engineering, Tsinghua University, Beijing 100084, China (e-mail: zcwang@tsinghua.edu.cn).


}

}



\maketitle

\begin{abstract}
Holographic multiple-input multiple-output (MIMO) systems constitute a promising technology in support of 
next-generation wireless communications, thus paving the way for a smart programmable radio environment. However, despite its significant potential, further fundamental issues remain to be addressed, such as the acquisition of accurate channel information. Indeed, the conventional angular-domain channel representation is no longer adequate for characterizing the sparsity inherent in holographic MIMO channels. To fill this knowledge gap, in this article, we conceive a decomposition and reconstruction (DeRe)-based framework for facilitating the estimation of sparse channels in holographic MIMOs. In particular, the channel parameters involved in the steering vector, namely the azimuth and elevation angles plus the distance (AED), are decomposed for independently constructing their own covariance matrices. Then, the acquisition of each parameter can be formulated as a compressive sensing (CS) problem by harnessing the covariance matrix associated with each individual parameter. We demonstrate that our solution exhibits an improved performance and imposes a reduced pilot overhead, despite its reduced complexity.
Finally, promising open research topics are highlighted to bridge the gap between the theory and the practical employment of holographic MIMO schemes.
\end{abstract}


\section{Introduction}
The paradigm shift from the fifth-generation (5G) to the sixth-generation (6G) systems requires new technologies that go beyond the current ones. As the ``crown jewel" of wireless technologies, massive multi-input and multi-output (mMIMO) solutions have been approved for inclusion in the 5G next radio (NR) standards, but their close relatives, such as large intelligent surfaces (LISs), reconfigurable intelligent surfaces (RISs), and holographic MIMOs, continue to evolve \cite{Holo-101,Holo-102,chen-vtm}. They are fabricated by programmable meta-materials based on low-cost, compact, light-weight, and power-efficient hardware architectures, aiming for constructing a smart programmable propagation environment.

Investigations of sophisticated RISs are in full swing both in academia and industry, indicating the substantial benefits of RISs when the line-of-sight (LoS) paths are obstructed. However, only marginal performance gains can be achieved in environments having strong LoS components, due to the double-hop fading effect experienced by the signal passing through the cascaded source-RIS-destination channels \cite{chen-vtm}. Furthermore, another drawback is its complex cascaded channel estimation, since excessive pilot overhead may be imposed on the practical systems. This erodes their benefits. Thus, we should maintain a neutral stance even though RISs have been touted as the ``technology star" for a while. In this context, a potential alternative technology is constituted by holographic MIMOs, which represents a natural extension/evolution of the current massive MIMOs. The holographic terminology means ``portray everything” in ancient Greek. The following subsections provide a brief overview of how it works, the state of research, and the road ahead.

\begin{figure*}[t]
	\centering
	\includegraphics[width=1.0\textwidth]{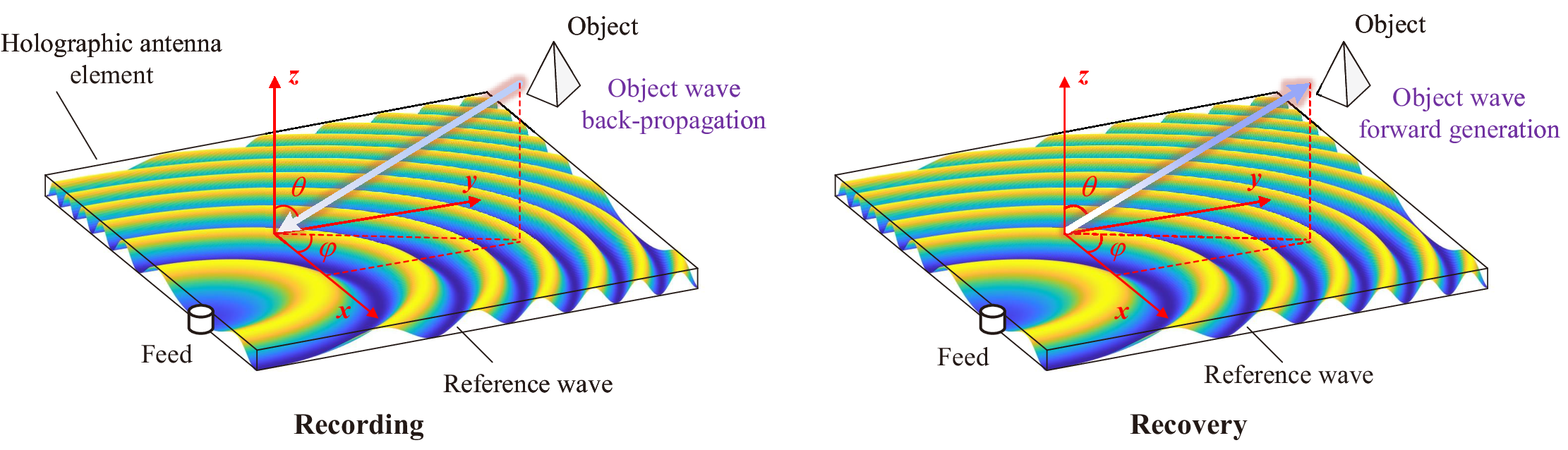}
	\caption{The holographic MIMO's beampattern is constructed by the superposition of the reference wave and the object wave~\cite{Holo-102}.} \label{antenna_element}
\end{figure*}

\subsection{Fundamentals}

The holographic MIMO surface is composed of a substrate with a few feeds and an extremely large number of antenna elements printed on its surface. The substrate functions as a waveguide, allowing electromagnetic waves (also termed as reference waves) to propagate along it. The feeds are used for relaying reference waves (information-carrying waves) coming from RF chains. The antenna elements are intended to establish various holograms, i.e., beampatterns, with the aid of their tuning mechanisms. 

To elaborate briefly, we interpret the operating principle of holographic MIMOs as a pair of recording and recovery processes \cite{Holo-105}. During the recording phase, a holographic beampattern can be constructed on the meta-surface for beneficially capturing the associated interferogram visualizing the interference between the reference wave and the object wave that back-propagates in the direction $(\theta, \varphi)$, as illustrated in Fig.~\ref{antenna_element}. During the recovery process, once the reference wave is stimulated and propagates along the surface, it is possible to generate a forward-propagating object wave in the $(\theta, \varphi)$-direction. Thanks to the integration of a theoretically infinite number of antenna elements, a holographic MIMO may be regarded as the approximate realization of a continuous aperture capable of constructing a sharp reconfigurable beampattern. Hence, this property leads to potential merits, including their thin profile, compact integration and fine angular resolution~\cite{Holo-101}. Its differentiating feature with respect to (w.r.t) RISs is that the element-feeds are implanted into the holographic meta-surface, whereas RISs have no internal feeds, hence, they can only work via passive reflection. Thus, holographic MIMOs can serve as active transceivers in demanding communication scenarios.



\subsection{State-of-the-art in Holographic MIMO Systems}

Although holographic MIMOs are still in their infancy, some impressive attempts have been undertaken. To be specific, as regards to channel modeling, a Fourier plane-wave series expansion of the channel response is provided in \cite{Holo-2}, which unveils the essence of electromagnetic propagation in arbitrary scattering and also works in the (radiative) near-field. Given the unavailability of the spatially correlated channel matrix, efficient channel estimation is conceived in \cite{Holo-4}, which exploits the array geometry to identify a subspace of reduced rank that covers the eigenspace of any spatial correlation matrix. Additionally, for attaining a high directional beam gain, the holographic transmit precoder (TPC) has to be carefully designed. Based upon this, the beampattern superposition of multiple users heralds a new kind of multiple access, termed as holographic-pattern division multiple access (HDMA) \cite{Holo-9}.


\subsection{Challenges and Motivations}

The complexity and pilot overhead are always of prime concern when it comes to channel estimation. In conventional mMIMO, the angular-domain channel estimation binds the magnitude of the pilot overhead to the number of channel paths, thus achieving considerable pilot savings. This, however, may not be applicable for holographic MIMO channels, since the structured sparsity inherent in the angular-domain representation fails to cope with the spherical wavefront in the presence of an extremely large antenna array, as it will become clear later. To exploit the specific nature of holographic MIMOs in the design of efficient channel estimation, the following issues require specific attention.

\textbf{Fact 1: Angular and distance power spread.}
Based on the classical planar-wavefront assumption, most channel estimation schemes exploit the sparsity of the angular domain for achieving pilot savings. However, in their holographic counterparts, the power scattering typically results in the proliferation of angles. Furthermore, the maximum scattered power of distance may not represent the accurate distance from the base station (BS) to the UE/scatterers.
This can be attributed to the fact that only a limited portion of the holographic array can sense the spherical electromagnetic waves radiated by a user or a scatterer. In this case, the power spectral density and the average received power fluctuate rapidly across the antenna array. Hence, the channel's output power corresponding to the paths having low average power may occasionally exceed that of its significant counterparts having high average power, thus eroding the detection accuracy. 


\textbf{Fact 2: Two dimensional (2D) angle plus distance search and pilot overhead.}
To quantify the sparseness of the channel matrix, a straightforward method is to carry out a joint sampling over the three-dimensional (3D) azimuth-elevation-distance (AED) parameters, which results in a potentially enormous search space and complexity, significantly undermining the efficiency of channel estimation. Thus, efficient techniques are required for searching through this 3D AED grid. Furthermore, the potentially excessive pilot overhead, which is theoretically proportional to the large number of antenna elements, can be reduced to a substantially lower order based exclusively on the number of channel paths.

%

For tackling the above critical challenges, we conceive a channel estimation framework, namely the decomposition-reconstruction (DeRe)-based technique, for holographic MIMOs by decomposing the 3D AED space into three one-dimensional (1D) spaces, followed by compressed reconstruction techniques. To elaborate, by constructing the 1D covariance matrix associated with the azimuth angle, the elevation angle, and the distance, respectively, the tightly coupled 3D AED parameters are decoupled to facilitate their independent estimation. 
Then, by exploiting the sparsity of the covariance matrix, the 3D AED parameters can be reconstructed as a 1D compressive sensing (CS) problem. We will demonstrate that by harnessing our proposed DeRe framework, the deleterious effects of power spread across scattered paths can be mitigated, and hence, the significant and insignificant paths may be confidently distinguished. Finally, the search (sampling) complexity can be beneficially reduced, because it becomes proportional to the sum - rather than to the product - of the number of azimuth, elevation, and distance sampling points.




\section{Electromagnetic Channel Characteristics}
The leap from mMIMO to holographic MIMO involves not only an increase in antenna scale, but also a fundamental shift in the structure of the electromagnetic propagation field. Exact and tractable wave propagation channel models are required for the practical design and performance evaluation of holographic MIMOs. 
In contrast to deterministic models requiring the numerical solution of Maxwell’s equations, stochastic models are capable of capturing the statistical characteristics of electromagnetic channels. The compact form-factor and the high number of antenna elements motivate new types of exploration in holographic MIMOs. Thus, to glean deeper insights, we commence by exploring several essential electromagnetic channel characteristics.

\subsection{Spatial Non-Stationarity}
When the number of holographic array elements becomes high, spatial non-stationary attributes tend to appear across the array. There may be different views of the propagation environment from different parts of the array, sensing the same channel paths with different powers, or even observing different channel paths \cite{XLM-102}. The non-stationary properties of arrays impose a departure from the commonly used correlated channel model. A cluster-based geometric channel model reflects more appropriately the cause of non-stationarity, with the expression of steering vectors being the major change in comparison to their traditional models. Firstly, the phase of each element should be modeled using spherical waves in the vicinity of the array, since the planar wave approximation is no longer viable. Secondly, the amplitude of each element fluctuates. This is attributed to the path loss along the array as well as to the interactions between clusters and blocks in the environment, because various segments of the spherical wavefront experience different propagation characteristics. Despite being site-specific, this modeling, which is dependent upon the relative position of the clusters and terminals w.r.t the array, may sustain high data rates in high-user-density scenarios.

\subsection{Spherical Wavefront}

The use of planar wavefront assumption is generally confined to the far-field regime routinely exploited for approximating the wavefronts as being locally planar throughout the whole array. The exploitation of a far-field model in the near-field regime may result in both magnitude and phase errors due to a non-negligible curvature of the incoming wavefronts. For example, at the Rayleigh distance of $R = 2{D^2}/\lambda $, a maximum phase error of $\pi / 8$ is encountered across an array of size $D$ at a wavelength $\lambda$ \cite{Holo-2,Fresnel}. In holographic MIMOs, the electromagnetically large arrays shift the electromagnetic operating domain from the far-field to the near-field, since they are characterized by a shorter far-field (Fraunhofer) distance than the conventional antenna arrays.

When modeling the spherical wavefront, a holographic MIMO draws its steering vector from an accurate spherical wave that concurrently incorporates the AED information. In this context, the phase of each element in the steering vector is a nonlinear function of the antenna index, but it also depends on the distance differences. Secondly, the scattered power usually results in multiple angles in the near field, as opposed to being concentrated in a single angle in the far-field region~\cite{XLM-1}. 
This power spread effect obscures our observation of the potential sparsity in near-field holographic MIMO channels.


\subsection{Rank-deficiency}
It is a common clich{\'e} to speak of rank-deficiency in massive MIMO, which refers to the inevitable spatial correlation among the channel realizations of the antennas.
As antennas are placed more densely, the ratio of the spatial correlation matrix rank to channel dimension tends to decline, which becomes even more pronounced in holographic MIMO channels. Furthermore, the establishment of the spatial correlation matrix and minimum mean square error (MMSE) estimates becomes quite arduous in a holographic MIMO system due to having thousands of antenna elements. Determining the MMSE estimates has a computational complexity order of $\mathcal{O} {\left( {N^3}\right) }$ due to the associated matrix inversion operation, and the pilot overhead is proportional to the number of antenna elements. These heavy computations and pilot overhead are not viable in practice. However, fortunately, the near-field channel is typically sparse due to the limited number of significant paths. By leveraging this, substantial pilot overhead savings can be achieved. 

Inspired by this promise, we explore the potential pilot savings, and the benefits of decoupling the 3D AED parameters to determine the estimates at reduced complexities. Next, let us move towards the DeRe framework for efficient channel estimation in holographic MIMOs by \textit{decomposing} the 3D AED parameters that are tightly coupled, followed by their respective 1D CS-based problems \textit{reconstructed}.

\section{Decomposition of 3D AED Parameters}

Prior to moving on, let us first revisit why existing  channel estimation schemes based on the far-field may become fruitless in the context of near-field holographic MIMO, thereby emphasizing the motivation behind this work. Subsequently, we will provide a detailed exposition of our motivation for the decomposition of the 3D AED parameters.

\subsection{Revisiting the Angular-Domain Representation}

\begin{figure}
	\subfigure[]{
		\begin{minipage}[t]{0.45\textwidth}
			\centering
			\includegraphics[width=\textwidth]{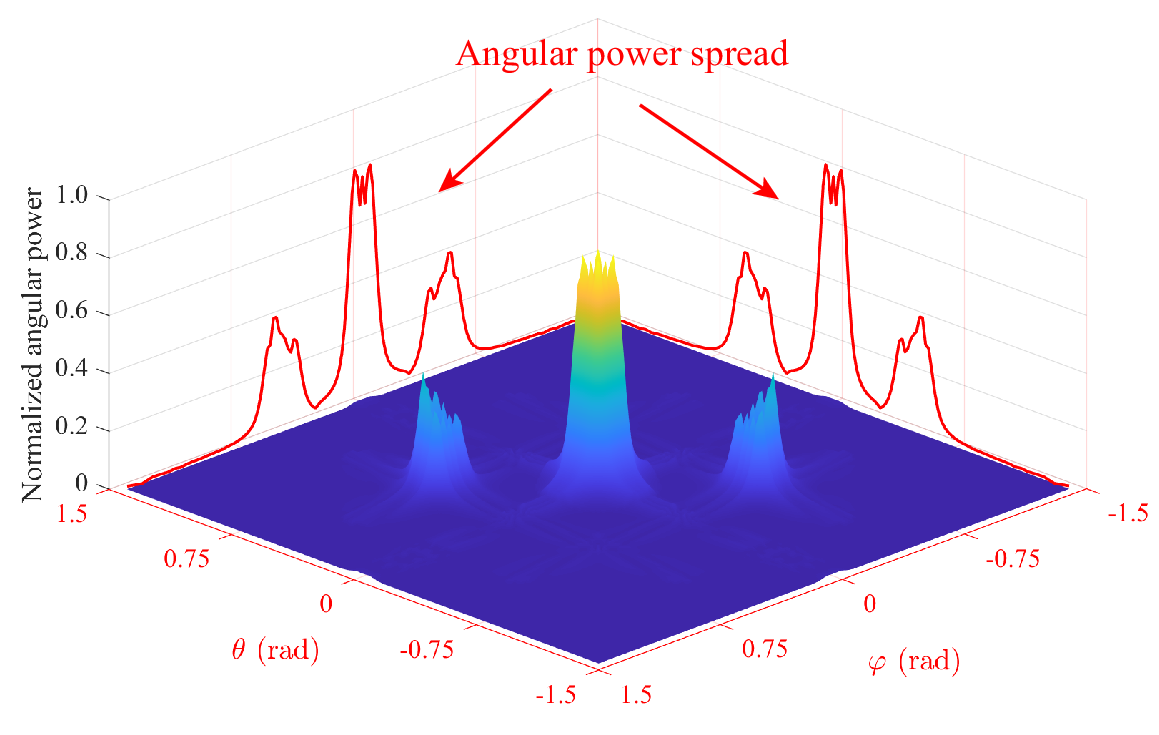}
			\label{3d_angular_spread}
	\end{minipage}}
	\subfigure[]{
		\begin{minipage}[t]{0.45\textwidth}
			\centering
			\includegraphics[width=\textwidth]{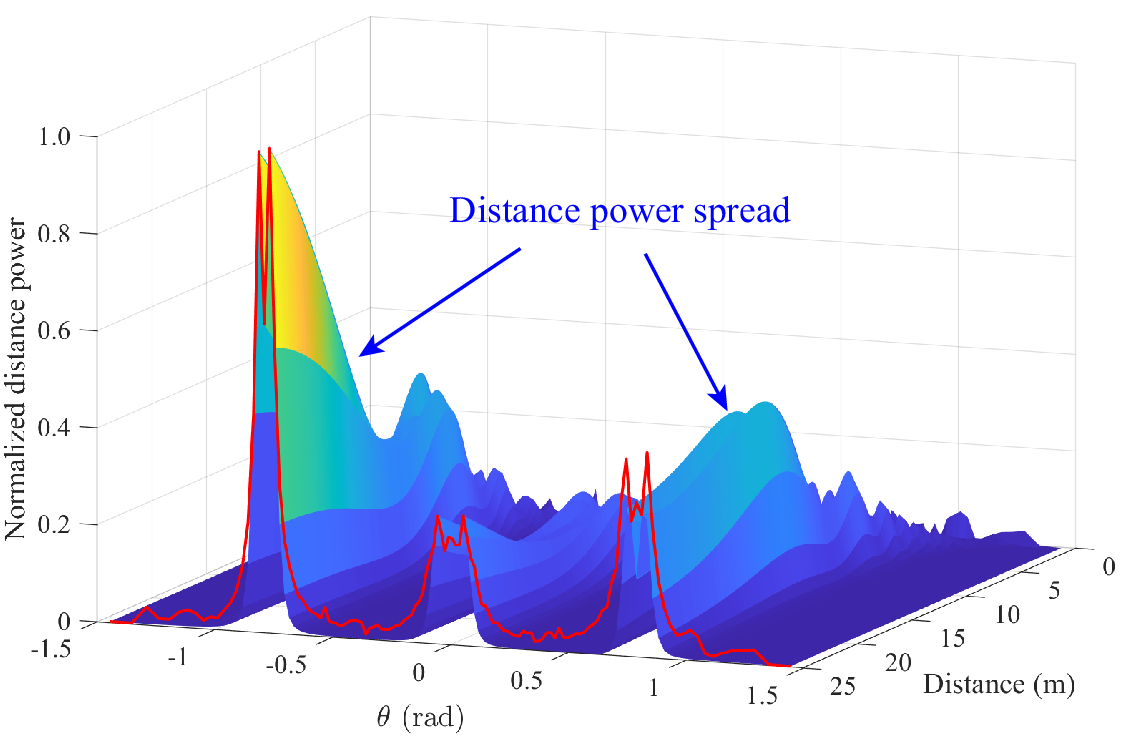}
			\label{3d_distance_spread}
	\end{minipage}}
	\caption{There are three scatterers in the propagation environment. (a) Angular power spread: the normalized power scattered results in the proliferation of a multitude angles. (b) Distance power spread: the normalized power scattered in relation to the distance experiences a fluctuating trend from peak to tough.}\label{challenge}
\end{figure}

The potential sparsity inherent in mMIMO channels can be exploited for efficient channel estimation with reduced complexity and pilot overhead. Specifically, given the limited number of scatterers that determine the signal’s propagation paths in the wireless environment, only a few pilots are required to detect the parameters (e.g., the azimuth-elevation angle pair $ \left( {\bm{\theta} ,\bm{\varphi} } \right)$) corresponding to the propagation paths, instead of acquiring all the  channel coefficients. These propagation paths are also known as significant paths. By harnessing the angular-domain representation, the power spectrum of each significant path can be exposed with the aid of the discrete Fourier transform (DFT) basis, as exemplified in \cite{AngDom-11}. In the context of a near-field holographic scenario, we are interested in examining the application of the angular-domain techniques for the observation of channel's sparsity. In pursuit of this, we  portray the angular-domain representation of the near-field holographic MIMO channels in Fig.~\ref{challenge}.

As it transpires from Fig.~\ref{3d_angular_spread}, the normalized scattered angular power typically results in the proliferation of angles, and fluctuates rapidly across the holographic antenna array. At a fixed distance, the true azimuth-elevation angle pair $ \left( {\bm{\theta} ,\bm{\varphi} } \right)$ cannot be precisely distinguished, as opposed to being concentrated on a few single peaks in the far-field scenario. By contrast, Fig.~\ref{3d_distance_spread} plots the scattered power vs. the distance in the  near-field holographic MIMO channel. Given that the position of the scatterer remains fixed in the propagation environment, we refer to the exact distance from the BS to the scatterer as the `genuine distance', where the scattered power may exhibit its peak value at this genuine distance. However, the use of the DFT basis for distance sampling might result in an erroneous conclusion, where the peak value of the scattered power is incorrectly taken to represent the distance between the BS and the UE/scatterers. As illustrated in Fig.~\ref{3d_distance_spread}, the power scattered vs. the distance reveals a fluctuating trend and the peak value of the scattered power may not indicate the genuine distance. The power may spread farther away towards an increased distance, termed as the distance power spread.

\subsection{Decomposition of 3D AED Parameters}

The steering vector of holographic MIMO encapsulates the extra distance traveled by the wave to arrive at an arbitrary antenna element w.r.t the geometric center of the UPA. The extra distance is also termed as the distance difference. By using the Fresnel approximation, the distance difference can be represented by the sum of a linear term and a quadratic term. Specifically, the linear term exhibits a monotonic function of angles and captures the planar-wave components within the near-field context, i.e., the direction information associated exclusively with the azimuth-elevation angle pair $ \left( {\bm{\theta} ,\bm{\varphi} } \right)$. By contrast, the quadratic term characterizes the curvature information of the spherical wavefront, involving the distance information. Then, we harness the popular concept of ``divide and conquer", which entails independently extracting the angular and the distance information from the holographic MIMO channel for estimating them one by one.

To ensure that the decomposed angles and distance still accurately represent the original channel characteristics, it is imperative to employ bespoke sparse functions ${\bf{W}}\left({\bm{\theta}} \right) $, ${\bf{W}}\left({\bm{\varphi}} \right) $, and ${\bf{W}}\left({\bf{r}} \right)$, specifically tailored for $\bm{\theta}$, $\bm{\varphi}$, and $\bf{r}$, respectively. The sparse function crafted aims for capturing the second-order statistics of each geometric parameter in the near-field holographic MIMO channel, functioning in essence as the covariance matrix. Within this sparse function, or equivalently, the covariance matrix, the non-zero diagonal entries reveal the channel's auto-correlations, while the associated non-zero off-diagonal entries indicate the cross-correlations. By harnessing these properties, the significant angles and distances can be accurately detected through the sparse functions constructed. Next, we elaborate on how to construct these sparse functions.

\begin{figure*}[t]
	\centering
	\includegraphics[width=0.95\textwidth]{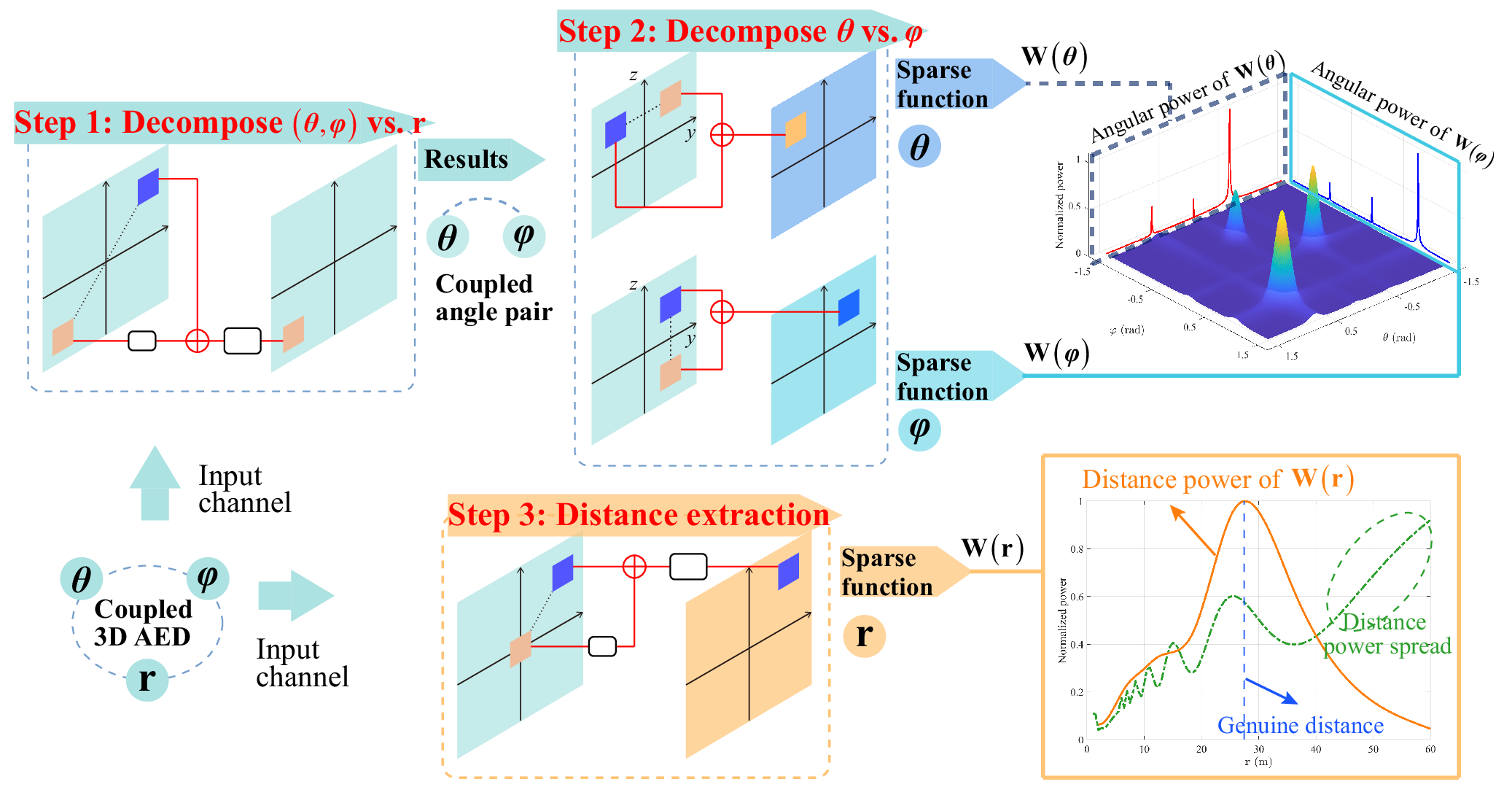}
	\caption{The parametric decomposition procedure for 3D AED parameters.} \label{decomp}
\end{figure*}

\subsubsection{\textbf{Step 1} - Decompose $ \left( {\bm{\theta} ,\bm{\varphi} } \right)$ vs. $\bf{r}$}

Given that each entry in the HMIMO channel matrix is directly linked to a specific antenna element index, we can carefully choose appropriate candidates from these entries to construct sparse functions. As illustrated in Fig. \ref{decomp}, by selecting indices that are symmetrically positioned relative to the origin and performing conjugate multiplication of their respective channel responses followed by taking the expectation, one may achieve a covariance entry that exclusively encapsulates the angle information $ \left( {\bm{\theta} ,\bm{\varphi} } \right)$.

\subsubsection{\textbf{Step 2} - Decompose $\bm{\theta}$ vs. $\bm{\varphi}$}

This step aims for decomposing the azimuth-elevation angle pair $ \left( {\bm{\theta} ,\bm{\varphi} } \right)$ by constructing their own sparse functions ${\bf{W}}\left({\bm{\theta}} \right) $ and ${\bf{W}}\left({\bm{\varphi}} \right) $, respectively. This renders an estimate of either $\bm{\theta}$ or $\bm{\varphi}$ each time, rather than both at once. Specifically, an entry of the sparse function ${\bf{W}}\left({\bm{\theta}} \right) $ is defined as the arithmetic mean of the covariance entry and of its counterpart having a converted index, as illustrated in Fig.~\ref{decomp}. It is intuitive that a pair of input indices of the sparse function ${\bf{W}}\left({\bm{\theta}} \right) $ are symmetric w.r.t the $z$-axis.

The $\bm{\varphi}$-based sparse function ${\bf{W}}\left({\bm{\varphi}} \right) $ can be constructed using a similar strategy relying on a pair of covariance entries with distinct indices. More precisely, an entry within the sparse function ${\bf{W}}\left({\bm{\varphi}} \right) $ is defined as the arithmetic mean between a given covariance entry and its counterpart having a converted index, in which the pair of input indices are symmetric w.r.t the $y$-axis.

The outcome of Step 2 is portrayed in Fig. \ref{decomp}. Having eliminated the impact of the distance, the sparsity of ${\bf{W}}\left({\bm{\theta}} \right) $ and ${\bf{W}}\left({\bm{\varphi}} \right) $ w.r.t the significant directions is only related to the azimuth-elevation angle pair in this propagation environment having three scatterers. Furthermore, the side projection allows for a clear observation of the scattered power associated with the azimuth angle $\bm{\theta}$ and elevation angle $\bm{\varphi}$ for each scatterer. 

\subsubsection{\textbf{Step 3} - Distance Extraction}

This step aims for constructing the distance sparse function ${\bf{W}}\left({\bf{r }} \right)$. Explicitly, as illustrated in Fig.~\ref{decomp}, an entry of ${\bf{W}}\left({\bf{r }} \right)$ can be obtained by selecting a pair of entries from the HMIMO channel matrix. For this specific pair of entries, one is arbitrarily selected while the other remains fixed. The index of the fixed entry is linked to the origin of a UPA. Subsequently, conjugate multiplication is performed on the pair, followed by taking the expectation. By exploiting the sparsity in the distance function, the scattered power w.r.t the genuine distance is prominent, as observed for the two peaks corresponding to each scatterer presented in Fig.~\ref{decomp}.

The results of Fig.~\ref{decomp} exhibit a pronounced contrast to the angular and distance power spreads illustrated in Fig.\ref{challenge}. Thus, the decomposition of the 3D AED parameters facilitates an efficient retrieval of these parameters from the sparse functions constructed.

\begin{figure*}[t]
	\centering
	\includegraphics[width=0.95\textwidth]{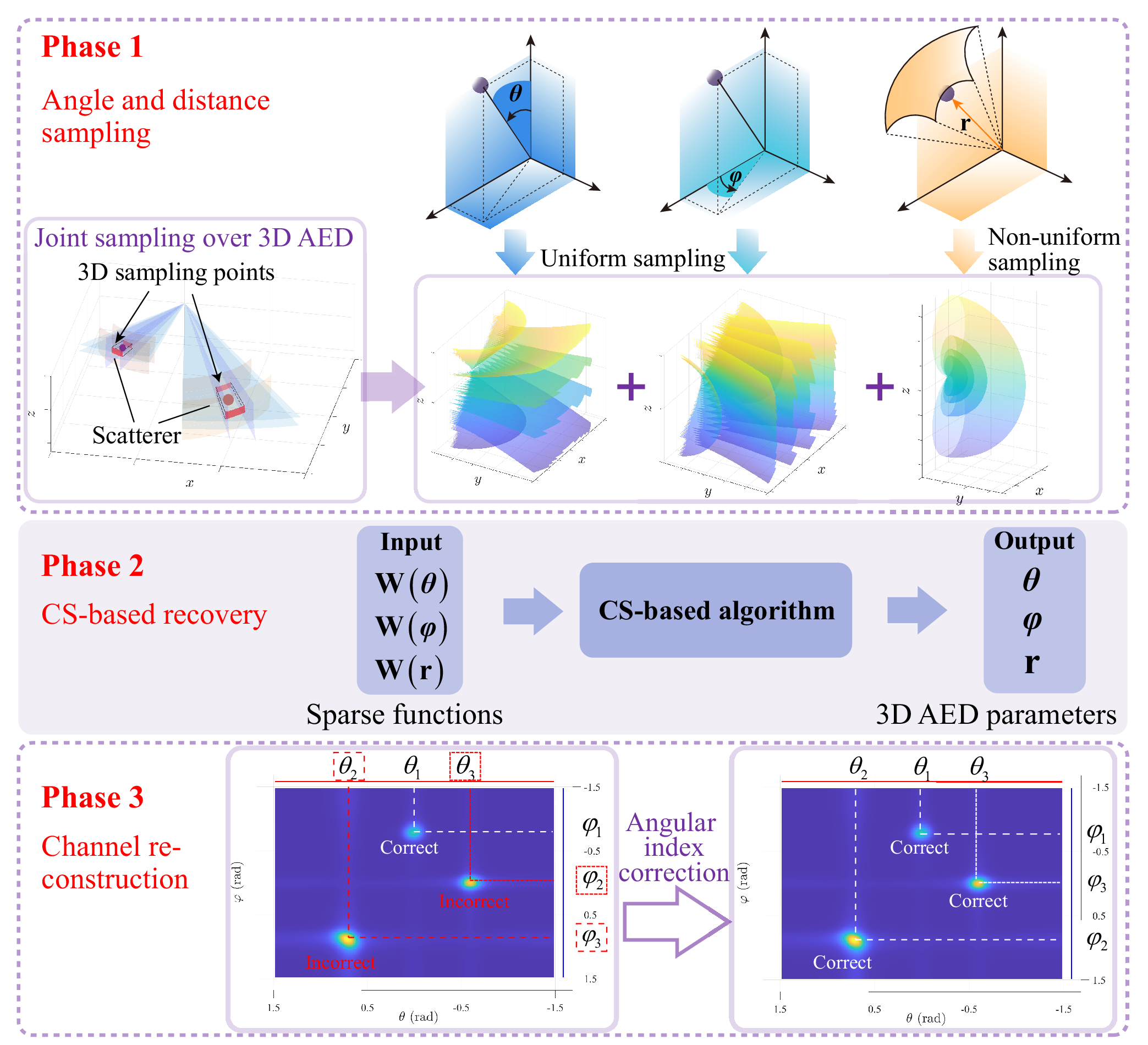}
	\caption{Sparse reconstruction of holographic MIMO channels.} \label{reconst}
\end{figure*}

\section{Reconstruction of Holographic MIMO Channels}
The goal of channel estimation is to reconstruct the 3D AED parameters for the precise recovery of holographic MIMO channels. A fundamental issue is to carry out efficient channel sampling of the 3D AED parameters. Following the CS framework~\cite{AngDom-11}, the dictionary matrix representing the angle and distance sampling has to satisfy stringent orthogonality for each of the two columns. Yet, the application of a conventional dictionary matrix, such as the DFT matrix, for jointly sampling the 3D AED parameters typically results in inaccuracies due to the associated parameter couplings. Additionally, its brute-forth nature results in a sampling complexity that is proportional to the product of the individual complexity of each parameter dimension.

Fortunately, throughout the preceding decomposition, we have converted the joint 3D estimate of $\bm{\theta}$, $\bm{\varphi}$, and $\bf{r}$ into their independent 1D estimates. This approach allows for the individual sampling of each parameter through its sparse function, facilitating the extraction of their respective significant power. In what follows, we commence with the implementation of effective sampling  versus the angle and distance using the sparse functions constructed, which is a critical step in harnessing the inherent sparsity. This enables the robust recovery of the 3D AED parameters and holographic MIMO channels, followed by demonstrating the performance attained.

\subsection{CS Based Reconstruction}
Again, each sparse function has sufficient channel information supporting for the recovery of 3D AED parameters. Below, we conceive a three-phase procedure for the sparse reconstruction of holographic MIMO channels.

\subsubsection{Phase 1 - Angle and Distance Sampling}

In sampling the angles $\bm{\theta}$ and $\bm{\varphi}$, the independence of ${\bf{W}}\left({\bm{\theta}} \right) $ and ${\bf{W}}\left({\bm{\varphi}} \right) $ allows for the uniform sampling of the azimuth and elevation angles by employing the classical DFT matrix in the angular domain \cite{Bayesian-0}. With the significant angles determined for distance sampling, we just need to discretize the distance along these identified directions. However, the inherent non-linearity of the distance-related terms characterized by the Fresnel approximation precludes the application of uniform sampling. The design of the dictionary matrix, therefore, has to be guided by the rule of minimizing the inter-column correlation to ensure optimal performance \cite{XLM-1}.

The appropriate sampling is achieved by the careful design of the dictionary matrix, aiming for extracting the relevant sparse functions. Then, each sparse function constructed can be structured in a standard CS form, represented by the product of a dictionary matrix and a sparse vector. 
This thus paves the way for efficiently recovering the 3D AED parameters at a reduced overhead with the aid of CS-based methods.

\subsubsection{Phase 2 - CS-Based Recovery}

Clearly, the channel estimation of holographic MIMO systems has now been now split into three independent CS-based parameter recovery problems w.r.t $\bm{\theta}$, $\bm{\varphi}$, and $\bf{r}$. Among the various solutions conceived for the classical parameter recovery problem, orthogonal matching pursuit (OMP) and variational Bayesian inference (VBI) stand out due to their superior performance \cite{chen-twc3}, as it will be evidenced in the subsequent discussions. We then obtain the sparse estimates of $\bm{\theta}$, $\bm{\varphi}$, and $\bf{r}$, respectively.

\begin{figure*}[t]
	\centering
	\includegraphics[width=0.95\textwidth]{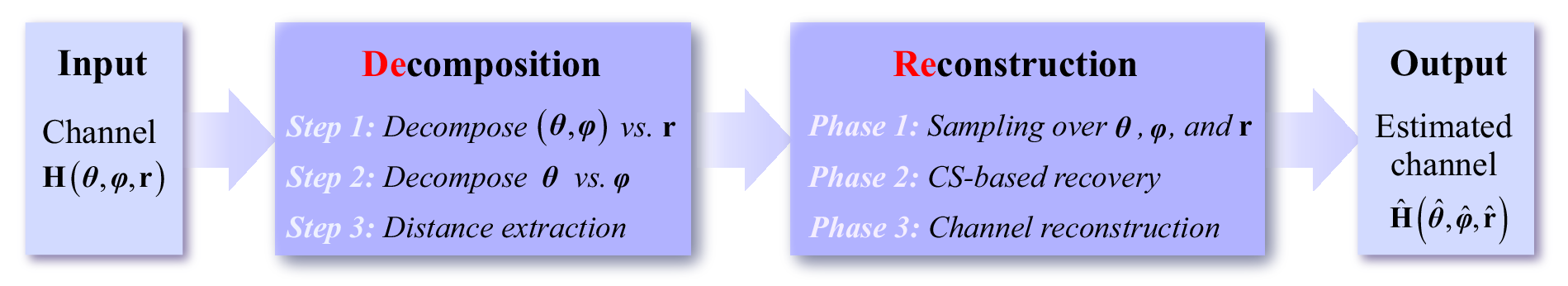}
	\caption{The proposed DeRe framework for holographic MIMO channel estimation.} \label{DeRe}
\end{figure*}

\subsubsection{Phase 3 -  Channel Reconstruction}

The outputs of the CS-based algorithm are $\bm{\theta}$, $\bm{\varphi}$, and $\bf{r}$.
When reconstructing the original holographic channel using the three independent information obtained, typically there is some angular index misinterpretation which significantly deteriorates the estimation performance. For example, consider a propagation environment where the significant paths are indexed by 1, 2, and 3. As illustrated in Fig.~\ref{reconst}, it is not given that the angle pair $\left( \theta_2, \varphi_3\right) $ is associated with path~2. By contrast, a more realistic guess is that $\left( \theta_2, \varphi_2\right) $ is the correct angle pair for path~2. Given that the distance sampling is carried out based on the specific directions determined by $\left( {\bm{\theta}, \bm{\varphi} } \right) $, there exists an exact match between the distance index and the angle pair index. This implies that the correction of the angle indices suffices for identifying a significant path.

For an accurate reconstruction of holographic MIMO channels, the idea of correcting the angular index misinterpretation is straightforward. Explicitly, this is based on identifying the two angles having the most similar  power, and matching their indices to the index of the path. Furthermore, since we have retrieved the index and amplitude (i.e., the complex gain) for $\bm{\theta}$ and $\bm{\varphi}$ using the CS technique in Step~2, correcting the angular index allows us to determine the complex gain for each path.
Therefore, we can achieve the accurate reconstruction of holographic MIMO channels.


\subsection{DeRe Framework}

Fig.~\ref{DeRe} illustrates the workflow of the proposed DeRe framework, which encompasses two core procedures: decomposition and reconstruction. Specifically, DeRe takes the original holographic MIMO channel as its input and performs parameter decomposition by constructing sparse functions for 3D AED parameters. Following this, robust recovery of these parameters across each dimension is achieved using CS-based techniques. Angular index correction is then employed for precisely identifying the indices of angle pairs with their respective significant path indices, ensuring accurate estimation of the holographic MIMO channel. Additionally, the DeRe framework is agnostic of the uplink vs. downlink regimes, offering flexible applicability across various estimation scenarios, as required.


\subsection{Performance and Discussion}

\begin{figure}[t]
	\centering
	\includegraphics[width=0.48\textwidth]{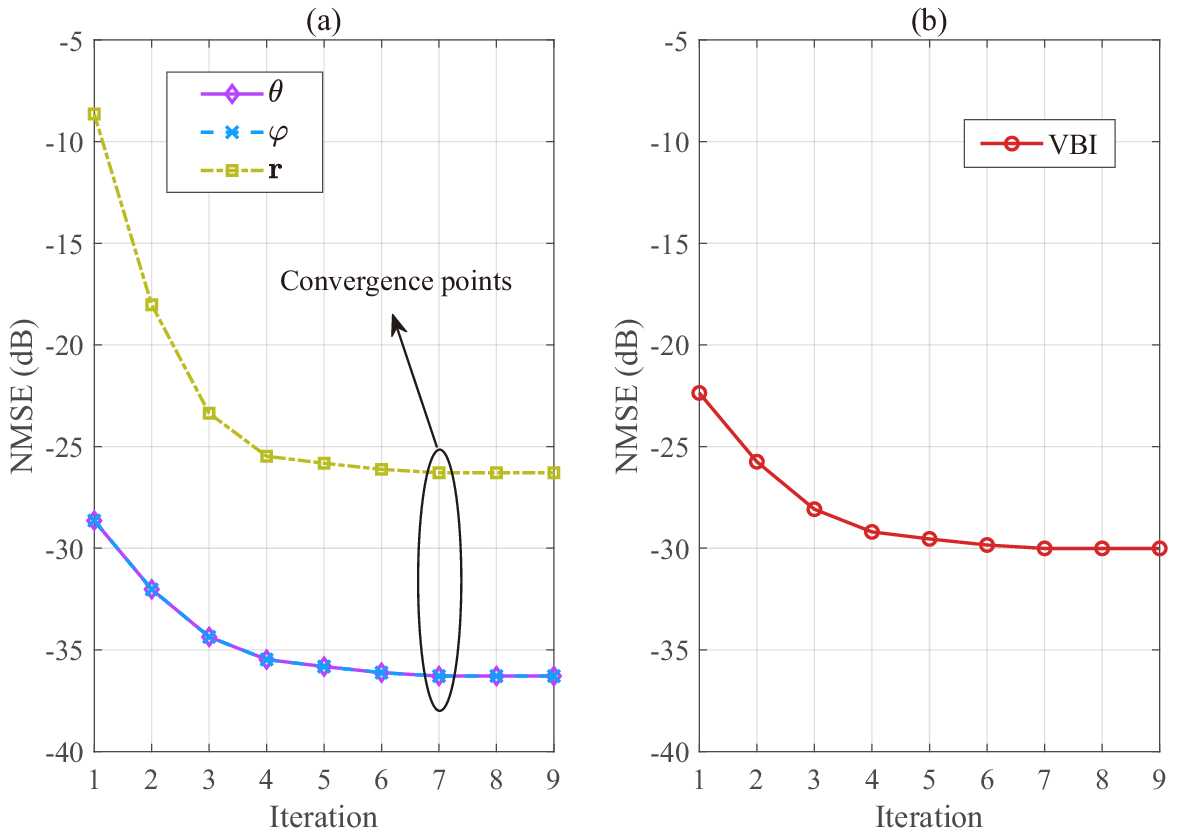}
	\caption{Convergence behavior of the VBI algorithm \cite{chen-twc3}. (a) NMSE performance of 3D AED parameters  $\bm{\theta}$, $\bm{\varphi}$, and $\bf{r}$. (b) NMSE performance of the holographic MIMO channel matrix.} \label{converg}
\end{figure}

\begin{figure}[t]
	\centering
	\includegraphics[width=0.48\textwidth]{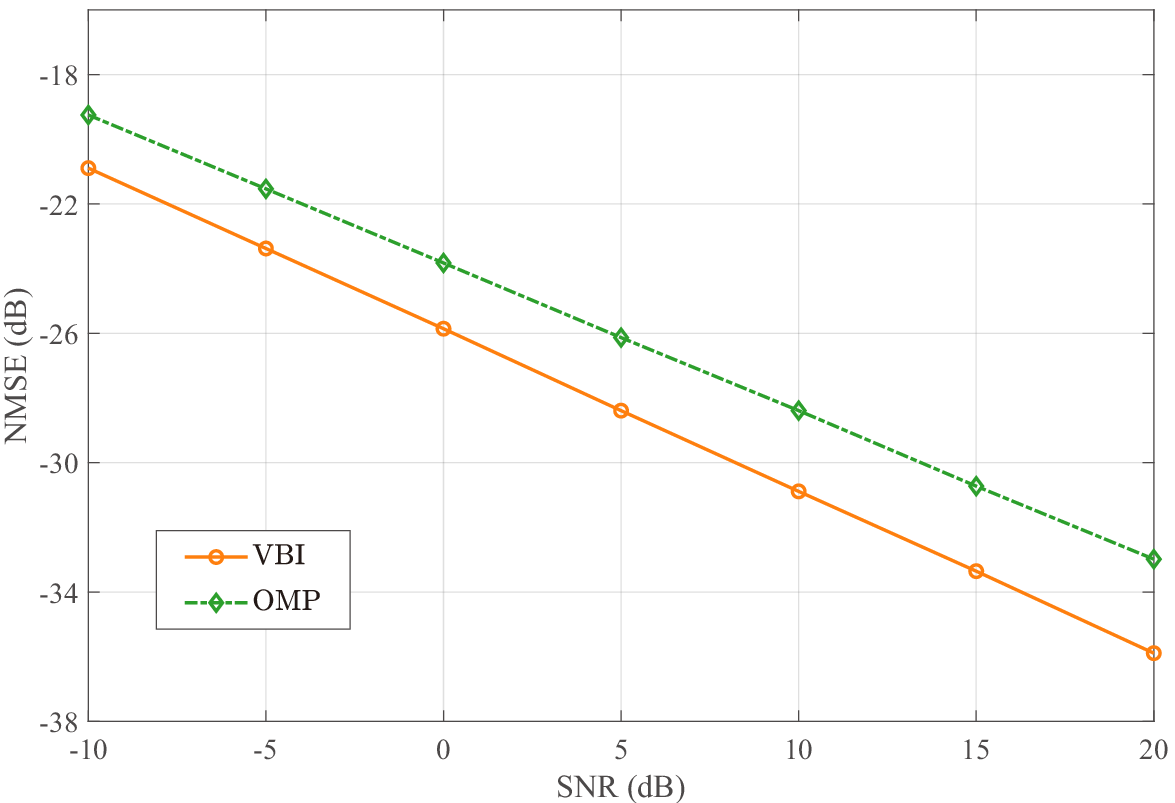}
	\caption{NMSE performance versus SNR. The antenna elements of the UPA are equipped with in the form of $129 \times 65$. The considered holographic MIMO system operates at 30~GHz. The number of paths is 3.} \label{NMSEvsSNR}
\end{figure}

Let us now characterize the convergence behavior of the VBI algorithm presented in \cite{chen-twc3} by plotting the normalized mean square error (NMSE) curves of the 3D AED parameters $\bm{\theta}$, $\bm{\varphi}$, and $\bf{r}$ in Fig.~\ref{converg}(a). By contrast, Fig.~\ref{converg}(b) shows the NMSE performance of the holographic MIMO matrix.  
A steep decline can be spotted in the first few iterations,
followed by a gradual settling at a value after around seven
iterations for both scenarios. The NMSE gap in accuracy
between $\bf{r}$ and $\bm{\theta}$ can be attributed to the presence of a residual correlation within the distance dictionary matrix. Despite this correlation, the VBI algorithm still exhibits favorable estimation performance. Additionally, Fig.~\ref{NMSEvsSNR} demonstrates the NMSE performance versus the signal-to-noise ratio (SNR) achieved by OMP \cite{XLM-1} and VBI \cite{chen-twc3} algorithms, respectively. It is observed that both algorithms reveal favorable estimation accuracy under the proposed DeRe paradigm.

Moving forward, let us examine the sampling complexity and pilot overhead imposed by the channel estimation procedure under the proposed DeRe framework. Phase~1 determines the sampling complexity of the reconstruction procedure. Due to the independent and parallel nature of the sampling process for $\bm{\theta}$, $\bm{\varphi}$, and $\bf{r}$, the original sampling complexity is relaxed to become proportional to the sum - rather than to the product - of the number of AED sampling points, hence significantly reducing the computational burden in practice. Additionally, Phase~2 determines the pilot overhead. For conventional estimation methods (such as least square) dispensing with our DeRe procedure, the pilot overhead is related to the number of holographic antenna elements. By contrast, in the presence of DeRe framework, the pilot overhead required for recovering each parameter is reduced to the order of the number of propagation paths, or potentially even lower \cite{Bayesian-0}.

\section{Future Directions}
Holographic MIMO schemes have substantial potential, but they are still in their infancy. Future research has to explore the implementation holographic MIMO systems, including the impact of estimation/feedback errors, hardware-imperfections, and practical holographic beampattern designs. Additionally, high-frequency millimeter wave and Terahertz systems, and a wider range of applications (such as radar, sensing, localization and so forth) using holographic MIMOs have to be explored in collaborative efforts. Below, we provide several promising topics for future exploration.

\textbf{Source Localization in Holographic MIMO Systems Using the DeRe Framework:} 
Viewed through the CS-based sparse recovery theory, both channel estimation and source localization share similar mathematical structures. As such, the DeRe framework can be employed for adaptively decomposing 3D AED parameters in the process of source localization in holographic MIMO systems.
However, the source localization is concentrated on the power aggregation of LoS paths emanating from different sources towards the same location~\cite{chen-jsac2}. Therefore, an appropriate sparsity model is required for distinguishing the LoS and NLoS paths, which is crucial for efficiently achieving accurate source localization in holographic MIMO systems.

\textbf{Integrating Artificial Intelligence (AI) with the DeRe Framework:} AI-empowered Radio Access Networks (RAN) have emerged as an important research topic and has been incorporated in 3GPP~TR~38.843 \cite{3GPP-AI}. These efforts highlight the trend towards utilizing AI for CSI acquisition in future wireless systems. Therefore, the DeRe framework developed in this article can be enhanced by AI. For example, AI could replace the CS-based methods in Phase~2 of Fig.~\ref{reconst} during the reconstruction procedure. Even more ambitiously, AI may learn the correlation among the 3D AED parameters, facilitating accurate reconstruction of holographic MIMO channels.

\textbf{Unified Far-Field and Near-Field Modeling and Signal Processing:}
In practical holographic MIMO systems, some users and scatterers may be located far away from the BS, whereas others may be near the BS, thus resulting in hybrid far- and near-field scenarios. Since the Rayleigh distance is proportional to the frequency, the near-field distance of ultra-broadband systems may fluctuate drastically across the entire bandwidth. Hence, the corresponding unified channel modeling and signal processing techniques deserve further research. Further investigation into the application of the DeRe framework in the unified representation of far-field and near-field is also an intriguing issue for holographic MIMO systems

\section{Concluding Remarks}

In this article, a DeRe-based framework was conceived for tackling a pair of critical challenges in holographic MIMO channel estimation. In particular, the 3D AED parameters were decoupled by establishing their associated covariance matrices. Then, the recovery of each parameter can be accomplished by formulating the corresponding CS estimation problem. As a result, the associated complexity was dramatically reduced. Furthermore, the power spread across scattered paths was mitigated, which resulted in an improved performance. Given that the investigation of holographic MIMO is still in its infancy, substantial future research is required in support of their evolution.


\ifCLASSOPTIONcaptionsoff
  \newpage
\fi

\bibliographystyle{IEEEtran}
\bibliography{ref_vtm}

\begin{IEEEbiography}[{\includegraphics[width=1in,height=1.25in,clip,keepaspectratio]{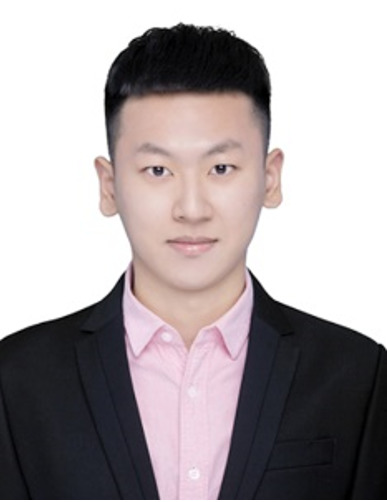}}]{Yuanbin Chen} (chen\_yuanbin@163.com) is currently pursuing the Ph.D. degree in information and communication systems with the School of Information and Communication Engineering, Beijing University of Posts and Telecommunications. He was honored with the National Scholarship in 2020 and 2022, and was awarded the second prize in the 2023 IEEE ComSoc Four-Minute-Thesis (4MT) Competition.
\end{IEEEbiography}

\begin{IEEEbiography}[{\includegraphics[width=1in,height=1.25in,clip,keepaspectratio]{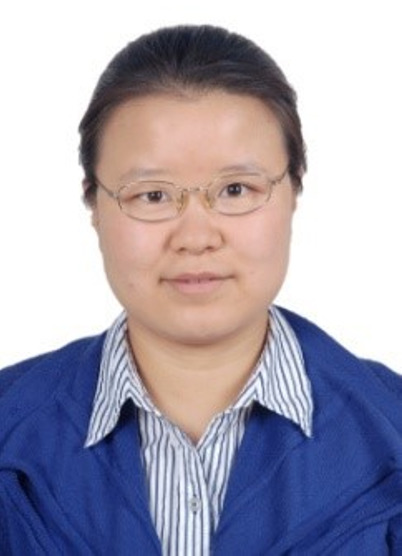}}]{Ying Wang} (wangying@bupt.edu.cn) is a  Professor with the School of Information and Communication Engineering, Beijing University of Posts and Telecommunications. Her research interests are in the area of the cooperative and cognitive systems, radio resource management, and mobility management in 5G systems.
\end{IEEEbiography}

\begin{IEEEbiography}[{\includegraphics[width=1in,height=1.25in,clip,keepaspectratio]{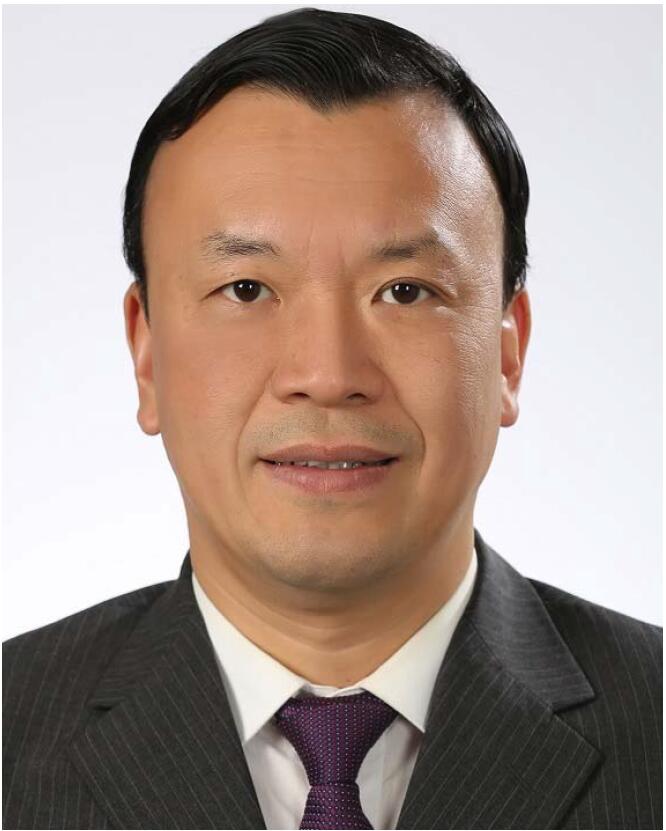}}]{Zhaocheng Wang}  (zcwang@tsinghua.edu.cn) received his B.S., M.S., and Ph.D. degrees from Tsinghua University in 1991, 1993, and 1996, respectively. From 1996 to 1997, he was a Post Doctoral Fellow with Nanyang Technological University, Singapore. From 1997 to 2009, he was a Research Engineer/Senior Engineer with OKI Techno Centre Pte. Ltd., Singapore. From 1999 to 2009, he was a Senior Engineer/Principal Engineer with Sony Deutschland GmbH, Germany. Since 2009, he has been a Professor with Department of Electronic Engineering, Tsinghua University. He was a recipient of IEEE Scott Helt Memorial Award, IET Premium Award, IEEE ComSoc Asia-Pacific Outstanding Paper Award and IEEE ComSoc Leonard G. Abraham Prize.
\end{IEEEbiography}

\begin{IEEEbiography}[{\includegraphics[width=1in,height=1.25in,clip,keepaspectratio]{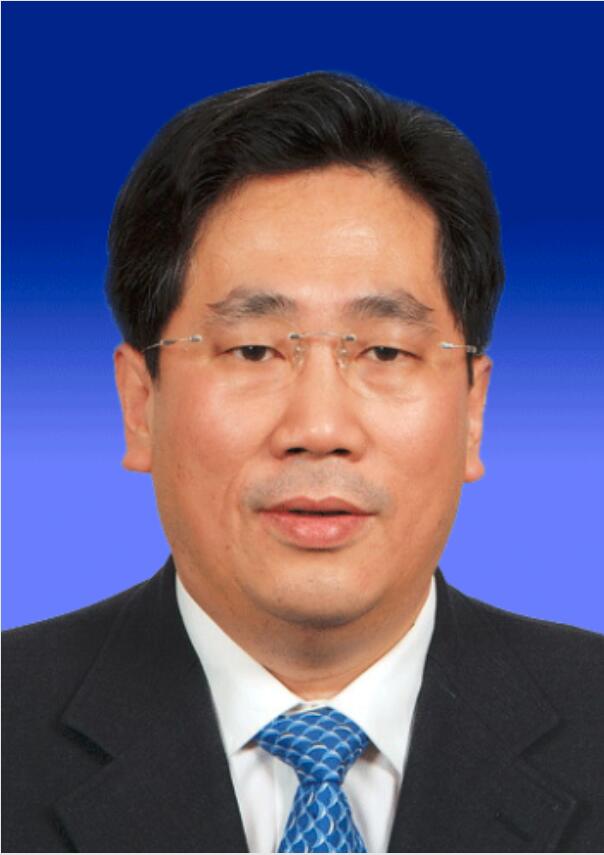}}]{Ping Zhang} (pzhang@bupt.edu.cn) is a Professor with the School of Information and Communication Engineering, Beijing University of Posts and Telecommunications, the Director of the State Key Laboratory of Networking and Switching Technology, a member of IMT-2020 (5G) Experts Panel, and a member of Experts Panel for China’s 6G Development. He served as a Chief Scientist of National Basic Research Program (973 Program), an Expert in information technology division of National High-Tech Research and Development Program (863 Program), and a member of Consultant Committee on International Cooperation of National Natural Science Foundation of China. His research interests mainly focus on wireless communications.
\end{IEEEbiography}

\end{document}